\documentstyle[11pt,aaspp4]{article}

\renewcommand{\plottwo}[2]{\epsscale{0.65} \plotone{#1} \\
   \plotone{#2} \epsscale{1.0}}

\begin{document}

\title{Starburst-like Dust Extinction in the Small Magellanic Cloud}

\author{Karl D. Gordon and Geoffrey C. Clayton}
\affil{Department of Physics \& Astronomy, Louisiana State University \\
   Baton Rouge, LA 70803 \\ email: (gordon,gclayton)@fenway.phys.lsu.edu }

\lefthead{Gordon \& Clayton}
\righthead{SMC Extinction}

\begin{abstract}
  The recent discovery that the UV dust extinction in starburst
galaxies is similar to that found in the Small Magellanic Cloud (SMC)
motivated us to re-investigate the ultraviolet (UV) extinction found
in the SMC.  We have been able to improve significantly on previous
studies by carefully choosing pairs of well matched reddened and
unreddened stars.  In addition, we benefited from the improved S/N of
the NEWSIPS {\it IUE} data and the larger sample of SMC stars now
available.  Searching the {\it IUE} Final Archive, we found only four
suitable early-type stars that were significantly reddened and had
well matched comparison stars.  The extinction for three of these
stars is remarkably similar. The curves are roughly linear with
$\lambda^{-1}$ and have no measurable 2175~\AA\ bump.  The fourth star
has an extinction curve with a significant 2175~\AA\ bump and weaker
far-UV extinction.  The dust along all four sightlines is thought to
be local to the SMC.  There is no significant Galactic foreground
component.  The first three stars lie in the SMC Bar and the
line-of-sight for each of them passes through regions of recent star
formation.  The fourth star belongs to the SMC Wing and its
line-of-sight passes though a much more quiescent region.  Thus, the
behavior of the dust extinction in the SMC supports a dependence of
dust properties on star formation activity.  However, other
environmental factors (such as galactic metallicity) must also be
important.  Dust in the 30 Dor region of the LMC, where much more
active star formation is present, does not share the extreme
extinction properties seen in SMC dust.
\end{abstract}


\section{Introduction}

  The interstellar dust in the Small Magellanic Cloud (SMC) has gained
new importance with the discovery that its ultraviolet (UV) extinction
is uniquely similar to extinction found in starburst galaxies
(\cite{cal94}; Gordon, Calzetti, \& Witt 1997).  Dust in the Galaxy
and the Large Magellanic Cloud does not show starburst galaxy-like
extinction.  Starburst galaxies are the only type of galaxy that has
been detected from low to high ($z > 2.5$) redshifts (\cite{kin93};
\cite{ste96}; \cite{low97}; \cite{tra97}; \cite{fra97}).  These
galaxies serve as an excellent probe of galaxy evolution through the
study of their star formation and metal enrichment rates at different
redshifts.  The derivation of these rates for high redshift galaxies
is sensitive to the adopted UV dust extinction (\cite{pet97};
\cite{mad97}).  Thus, understanding the physical processes responsible
for producing the starburst-like dust seen in the SMC is important to
the modeling of starburst galaxies.

  In addition, the study of the UV extinction curve in Local Group
galaxies promises to help in determining the sizes, shapes, and
materials which make up dust grains.  Ultraviolet extinction curves
have been determined in the Milky Way, the Large Magellanic Cloud
(LMC), the SMC, and (tentatively) M31.  The extinction curves in these
four galaxies paint a complex picture of the environmental dependence
of dust properties.  In the Milky Way, Cardelli, Clayton, \& Mathis
(1989, hereafter \cite{car89}) found that the infrared to UV
extinction curves can be described fairly well by a relationship which
depends on only one parameter, $R_V = A_V/E(B-V)$ which is a measure
of the overall dust grain size.  There are significant deviations from
the \cite{car89} relation in different Galactic environments
(\cite{mat92}).  In the LMC, the UV extinction curves show a distinctly
different behavior between the 30~Dor region (a mini-starburst
[\cite{wal91}]) and the rest of the LMC (\cite{cla85}; \cite{fit85},
1986).  The 2175~\AA\ bump is weaker and the far-UV rise is stronger
in the 30~Dor region than in the rest of the LMC which have strengths
similar to the average Galactic extinction curve.  In the SMC,
the average extinction curve is characterized by a roughly linear rise
(versus $\lambda^{-1}$) increasing toward shorter wavelengths without
a 2175~\AA\ bump (\cite{pre84}; \cite{tho88}).  Yet, there is one
sightline which has an extinction curve with a significant 2175~\AA\
bump (\cite{leq82}).  In M31, the extinction curve is consistent with
that of the average Galactic extinction within the associated
uncertainties, although the 2175~\AA\ bump may be weak (\cite{bia96}).
The complex behavior in these four galaxies implies that the physical
properties of dust grains may be dependent on a multitude of
environmental parameters.  Two of these are metallicity and star
formation activity both of which may affect the overall composition
and size distribution of dust grains (\cite{cla85};
\cite{fit85}; \cite{car89}; \cite{gor97}).

\section{Previous Work \label{sec_previous}}

  There have been a number of papers based in part or entirely on the
derivation of SMC UV extinction curves (\cite{leq82}; \cite{pre84};
\cite{tho88}; \cite{rod97}).  These studies used the pair method to
derive extinction curves (see \S\ref{sec_calc}).  The extinction
curves were derived by comparing a reddened SMC star to an unreddened
SMC star or (more commonly) a group of unreddened SMC stars.  Using
unreddened stars of the same temperature and luminosity as the
reddened stars is crucial to determining accurate extinction curves.
All of the previous work suffered from poor spectral matching between
reddened and unreddened stars due in part to the paucity of accurate
optical and/or UV 2D spectral types.  Recent work has provided 2D
spectral types in both the optical (Garmany, Conti, \& Massey 1987;
\cite{mas95}; \cite{len97}) and UV (\cite{neu97}).  The previous
studies also suffered from low signal-to-noise spectra due to the
faintness of the stars in the SMC.  The signal-to-noise ratio and
absolute calibration of all the spectra taken by the International
Ultraviolet Explorer ({\it IUE}) have been greatly improved through
the NEWSIPS reduction routines (\cite{nic94}).  Also, there is now a
larger sample of SMC stars observed with {\it IUE} than were available
a decade or more ago when the previous studies were done. Thus, the UV
extinction curves in the SMC can be greatly improved using star pairs
that are better spectral matches and have higher S/N spectra.  We list
the reddened stars used in previous works in
Table~\ref{table_prev_work} along with comments on the quality of the
derived extinction curves.

\begin{deluxetable}{ccccl}
\tablewidth{0pt}
\tablecaption{Previous Extinction Work \label{table_prev_work}}
\tablehead{\colhead{AZV\tablenotemark{a}} & 
           \colhead{SK\tablenotemark{b}} &
           \colhead{ref\tablenotemark{c}} & 
           \colhead{quality\tablenotemark{d}} &
           \colhead{comments} }
\startdata
18\tablenotemark{e} &  13 & P84,T88 & good & good comparison \nl
20  &  14 & R97 & bad & spectral type too late (A0Ia) \nl
56  &  31 & T88 & bad & too bright for reddened star \nl
126 &     & R97 & bad & unreddened star \nl
211 &  74 & R97 & bad & spectral type too late (A0Ia) \nl
373 & 119 & T88 & bad & very low $E(B-V)$ \nl
398\tablenotemark{e} &     & P84,R97 & good & good comparisons \nl
456\tablenotemark{e} & 143 & L82,R97 & poor & spectral mismatch \nl
    & 191 & P84,T88 & bad & unreddened star \nl
\enddata
\tablenotetext{a}{\cite{azz75}; \cite{azz79}; 1982}
\tablenotetext{b}{\cite{san68}, 1969}
\tablenotetext{c}{L82 = \cite{leq82}; P84 = \cite{pre84}; T88 =
   \cite{tho88}; R97 = \cite{rod97}} 
\tablenotetext{d}{The quality of the extinction curve was good
(reddened star and at least one good comparison star), poor (reddened
star and no good comparison stars), or bad (star unsuitable for
extinction curve work)}
\tablenotetext{e}{Used in this study}
\end{deluxetable}

  Most of the previous work has been done by three groups, one based
in France, one in England, and one in Brazil.  The final results of
the French group are presented in Lequeux et al.\ (1982) and Pr\'evot
et al.\ (1984) which supersede earlier work (\cite{roc81};
\cite{leq84}).  In Pr\'evot et al. (1984), this group determined the
best (and often used) average UV extinction curve for the SMC from
only {\it three} reddened stars (AZV~18, AZV~398, \& SK~191).  One
drawback to this paper is that the individual extinction curves are
not shown, just the average which makes it difficult to assess the
accuracy of the individual curves.  The star, SK~191, is an
essentially unreddened star, making its extinction curve useless.  In
addition, Pr\'evot et al.\ (1984) derived extinction curves for two
other stars, but excluded them as having ``anomalous'' extinction
curves.  These other two stars were AZV~393 (\cite{leq84}) and AZV~456
(\cite{leq82}).  AZV~393 is an unreddened star with a UV spectral type
of B3~Ia (\cite{neu97}).  The extinction curve for AZV~456 is real and
significantly different from the average SMC extinction (Lequeux et
al.\ 1982). This is the only curve in the SMC which shows the presence
of the 2175~\AA\ bump.  However, the extinction curve, derived by this
group for AZV~456, suffers from a spectral mismatch which shows up
in an asymmetric 2175~\AA\ bump.  This is due to a mismatch between
the reddened and comparison stars in the \ion{Fe}{3} lines around 2000
\AA\ which are luminosity and temperature sensitive (\cite{car92};
\cite{neu97}).  The validity of the Pr\'evot et al.\ (1984) average
extinction curve is called into question through their inclusion of a
highly uncertain extinction curve (SK~191) and their exclusion of the
``anomalous'' extinction curve of AZV~456.

  The final results of the English group are contained in Thompson et
al.\ (1988) which supersedes earlier work (\cite{nan82};
\cite{bro83}).  In Thompson et al.\ (1988) a sample of four reddened
stars (AZV~18, 56, 373, \& SK~191), when compared to the unreddened
star AZV~264, produced a UV extinction curve consistent with the
Pr\'evot et al. (1984) average curve.  This was done instead of
calculating individual extinction curves as all of the reddened stars,
except for AZV~18, are only lightly reddened ($\Delta (B-V) < 0.13$)
making their extinction curves highly uncertain.

  The Brazilian group's work is presented in Rodrigues et al.\ (1997).
They calculate UV extinction curves for five stars -- AZV~20, 126,
211, 398, \& 456.  Two of these stars, AZV 20 \& 211, have A0~Ia
spectral types.  Stars with spectral types later than about B5 are
normally not used for UV extinction work.  This is due to their lower
UV fluxes and the rapid change in their intrinsic spectra as a
function of spectral type (\cite{rod97}).  These problems resulted in
extinction curves for these two stars (shown in Fig.~2 of Rodrigues et
al.\ [1997]) which are very noisy (AZV~20) or have odd changes of
slope (AZV~211).  The extinction curve for AZV~126 presented by
Rodrigues et al.\ (1997) has a very shallow slope which is a result of
comparing the unreddened star (AZV~126) to other unreddened stars of
earlier spectral types.  The UV spectral type of AZV~126 is B1~II and
the UV spectral types of the comparison stars, AZV~61, 317, and 454,
are O5~III, O7~Ia, and O9~V,respectively (\cite{neu97}).  The
extinction curves for AZV~398 and 456 presented by Rodrigues et al.\
(1997) reproduce the work of Pr\'evot et al.\ (1984).

  From a careful analysis of the previous work on the extinction
curves in the SMC, only three reddened stars (AZV~18, 398, \& 456)
emerge as good candidates for extinction curve work.

\section{Extinction Curves \label{sec_curves}}

   We collected all the stars with spectral types between O9 and B3 in
the SMC with {\it IUE} low dispersion spectra in order to have the
largest possible sample of reddened and unreddened stars.  Optical and
UV spectral types were taken from the literature (\cite{bou85};
\cite{gar87}; \cite{mas95}; \cite{len97}; \cite{neu97}).  From this
list, candidate reddened stars were identified as having red $(B - V)$
colors when compared to other stars with similar optical and UV
spectral types.

  For each candidate reddened star, we attempted to identify a
comparison star which satisfied the three Fitzpatrick criteria
(\cite{fit85}).  In addition, we required the $\Delta (B-V)$ between
the reddened and comparison star to be greater than 0.15.  The first
Fitzpatrick criterion requires that $\Delta (U-B)/\Delta (B-V)$ has a
value appropriate for reddening due to dust.  The average value of
$\Delta (U-B)/\Delta (B-V)$ for the SMC is $0.81 \pm 0.11$
(\cite{bou85}). The second Fitzpatrick criterion requires that $\Delta
V_o$, which is the difference between the dereddened $V$ magnitude of
the reddened star and the $V$ magnitude of the comparison star, is
less than 0.8 magnitudes.  This criterion insures the luminosities of
the two stars are comparable as all stars in the SMC are located at
approximately the same distance.  The reddened star's $V$ magnitude
was dereddened using $R_V = 2.72$ (\cite{bou85}) and the $\Delta
(B-V)$ between the reddened and comparison stars.  The third
Fitzpatrick criterion requires a good luminosity and temperature match
(i.e.\ UV spectral type) between the detailed spectra of the reddened
and comparison stars. The errors in the the UV spectral types are
$\sim$1 temperature subclass and $\sim$1 luminosity class
(\cite{neu97}).

  We applied these criteria to the candidate reddened stars and found
only one star, AZV~214, in addition to the three stars identified in
\S2, which satisfied all three of the Fitzpatrick criteria.  Thus, we
are left with a small sample of 4 reddened stars with which we can
study the UV extinction in the SMC.  Table~\ref{table_stellar_data}
lists the spectral and photometric data for the four reddened and four
comparison stars.  The UV spectral types are from Neubig \& Bruhweiler
(1997).  The optical and infrared photometry was taken from Bouchet et
al.\ (1985), except for AZV~70 which was taken from Ardeberg \&
Maurice (1977).  The uncertainties in the optical and infrared
photometry are 0.02, 0.045, 0.02, 0.028, 0.032, and 0.027 for $V$, $(U
- V)$, $(B - V)$, $(J - V)$, $(H - V)$, and $(K - V)$, respectively
(\cite{ard77}; \cite{bou85}; \cite{bou97}).
Table~\ref{table_fitz_criteria} displays the $\Delta (B-V)$, $\Delta
(U-B)/\Delta (B-V)$, and $\Delta V_o$ values for the reddened and
comparison pairs.

\begin{deluxetable}{cccllccccccc}
\small
\tablewidth{0pt}
\tablecaption{Stellar Data \label{table_stellar_data}}
\tablehead{ & & & \colhead{UV} & \colhead{Optical} & & & & & & & \\
           \colhead{type\tablenotemark{a}} & \colhead{AZV} & \colhead{SK} & 
           \colhead{sp.\ type} & \colhead{sp.\ type} &
           \colhead{ref\tablenotemark{b}} & \colhead{V} & \colhead{(U-V)} &
           \colhead{(B-V)} & \colhead{(J-V)} & \colhead{(H-V)} &
           \colhead{(K-V)} }
\startdata
r & 18 & 13 & B3 Ia & B2 Ia & L97 & 12.40 & -0.75 & 0.01 & -0.03 & -0.05 & 
   -0.08 \nl
c & 462 & 145 & B2 Ia & B1.5 Ia & L97 & 12.54 & -1.04 & -0.14 & 0.35 & 0.40 & 
   0.41 \nl
\tablevspace{5pt}
r & 214 & & B2 Ia & B3 Iab & G87 & 13.35 & -0.72 & 0.05 & 0.02 & 0.00 & 
   -0.03 \nl
c & 380 & 120 & B1 1a: & B0.5: & B85 & 13.50 & -1.00 & -0.11 & 0.27 & 0.29 & 
   0.37 \nl
\tablevspace{5pt}
r & 398 & & O9 Ia: & O9.7 Ia & B85 & 13.86 & -0.70 & 0.08 & -0.22 & -0.25 & 
   -0.26 \nl
c & 289 & 103 & O9 Ia & B0 I & B85 & 12.37 & -1.09 & -0.15 & 0.24 & 0.29 &
   0.34 \nl
\tablevspace{5pt}
r & 456 & 143 & O8 II & O9.7 Ib & B85 & 12.90 & -0.69 & 0.07 & -0.08 & -0.10 &
   -0.10 \nl
c & 70 & 35 & O9 Ia & O9.5 Iw & H83 & 12.39 & -1.13 & -0.16 & \nodata
   & \nodata & \nodata \nl
\enddata
\tablenotetext{a}{r = reddened star, c = comparison star}
\tablenotetext{b}{H83 = \cite{hum83}, B85 = \cite{bou85}, 
   G87 = \cite{gar87}, L97 = \cite{len97}}
\end{deluxetable}

\begin{deluxetable}{llccr}
\tablewidth{0pt}
\tablecaption{Fitzpatrick Criteria \label{table_fitz_criteria}}
\tablehead{\multicolumn{1}{c}{reddened} & \multicolumn{1}{c}{comparison} & 
           \colhead{$\Delta (B-V)$} & 
           \colhead{$\frac{\Delta (U-B)}{\Delta (B-V)}$} & 
           \multicolumn{1}{c}{$\Delta V_o$} }
\startdata
AZV 18  & AZV 462 & $0.15 \pm 0.03$ & $0.93 \pm 0.42$ & 
   $-0.55 \pm 0.09$ \nl
AZV 214 & AZV 380 & $0.16 \pm 0.03$ & $0.75 \pm 0.31$ & 
   $0.59 \pm 0.09$ \nl
AZV 398 & AZV 289 & $0.23 \pm 0.03$ & $0.70 \pm 0.26$ & 
   $0.86 \pm 0.10$ \nl
AZV 456 & AZV 70  & $0.24 \pm 0.03$ & $0.75 \pm 0.23$ & 
   $-0.13 \pm 0.10$ \nl
\enddata
\end{deluxetable}

\subsection{Calculation of Extinction Curves \label{sec_calc}}

  We calculated extinction curves using the standard pair method
(\cite{mas83}) which uses a reddened star and an appropriately
chosen unreddened comparison star.  The extinction curves were
calculated using
\begin{equation}
E(\lambda) = \frac{\Delta (\lambda - V)}{\Delta (B - V)} = 
   \frac{m(\lambda - V)_r - m(\lambda - V)_c}{(B - V)_r - (B - V)_c}
\end{equation}
where the subscripts $r$ and $c$ refer to the reddened and comparison
stars, respectively.  Individual short (SWP) and long (LWR and LWP) IUE
spectra were coadded resulting in a single spectrum extending from
1300 \AA\ to 3200 \AA.  The spectra were coadded using the
nearest neighbor method with bad points (as defined by the NEWSIPS
quality vector) excluded.  The individual spectra were weighted by
their exposure times.  The resulting long and short wavelength
spectra were binned to the instrumental resolution of $\sim$5 \AA.
The two spectra were then merged at the maximum wavelength in 
the short wavelength spectrum.  Table~\ref{table_iue} tabulates the
{\it IUE} spectra we used.

\begin{deluxetable}{rl}
\tablewidth{0pt}
\tablecaption{{\it IUE} Data\label{table_iue}}
\tablehead{\colhead{AZV} & \colhead{{\it IUE} Spectra} }
\startdata
18  & SWP 8295/10321/18014/18015 \nl
    & LWR 7241/14207 \nl
70  & SWP 16621/18830--LWP 12387 \nl
214 & SWP 22372--LWR 17263 \nl
289 & SWP 16049--LWR 12345 \nl
380 & SWP 10319--LWR17265 \nl
398 & SWP 18911/22361--LWR 14963 \nl
456 & SWP 16051/45199 \nl
    & LWR 12347--LWP 23556 \nl
462 & SWP 10316--LWR 17259 \nl
\enddata
\end{deluxetable}

  Uncertainties in the extinction curves were calculated using the
method of Massa, Savage, \& Fitzpatrick (1983) and Cardelli, Sembach,
\& Mathis (1992).  It is no longer necessary to estimate the
uncertainties in the {\it IUE} fluxes as these are calculated in the
NEWSIPS reduction.  The uncertainties in the extinction curve were
calculated using
\begin{equation}
\label{eq_main_err}
\sigma[E(\lambda)] = E(\lambda) \sqrt{ \left( 
   \frac{\sigma[\Delta (\lambda - V)]}{\Delta (\lambda - V)} \right)^2
   + \left( \frac{\sigma[\Delta (B - V)]}{\Delta (B - V)} \right)^2 }
\end{equation}
where
\begin{equation}
\sigma[\Delta (B - V)] = \sqrt{ \sigma[(B - V)_r]^2 + \sigma[(B - V)_c]^2 },
\end{equation}
\begin{equation}
\sigma[\Delta (\lambda - V)] = \sqrt{ \sigma[m(\lambda)_r]^2 +
   \sigma[V_r]^2 + \sigma[m(\lambda)_c]^2 + \sigma[V_c]^2 },
\end{equation}
\begin{equation}
\sigma[m(\lambda)] = \frac{-2.5}{2} \log \left( 
   \frac{F(\lambda) - \sigma[F(\lambda)]} {F(\lambda) +
    	\sigma[F(\lambda)]} \right),
\end{equation}
\begin{equation}
\label{eq_sig_fl}
\sigma[F(\lambda)] = \sqrt{ \sigma[F(\lambda)_{\rm NEWSIPS}]^2 +
   [\sigma_{\rm repeat} F(\lambda)]^2 };
\end{equation}
$F(\lambda)$ is the {\it IUE} measured flux; $\sigma[F(\lambda)_{\rm
NEWSIPS}]$ is the uncertainty NEWSIPS calculates; $\sigma_{\rm repeat}
\sim 0.05$\% is the estimated uncertainty in the relative {\it IUE}
calibration (the difference between two spectra of the same object
taken by {\it IUE} [\cite{gar94}]).  We have not included an error
term for temperature or luminosity mismatch as they are not easily
quantifiable.  Also, Cardelli et al.\ (1992) have shown that these
uncertainties are smaller than those calculated using
equation~\ref{eq_main_err}.  The uncertainties (as calculated in
equation~\ref{eq_main_err}) in the UV extinction curves can be divided
into two parts.  One part arises from random uncertainties in the UV
data and the magnitude of this uncertainty can be reduced by binning
the UV data.  The other part arises from random uncertainties in the
optical photometry and relative calibration of {\it IUE} which cannot
be reduced by binning the UV data.  Thus, a lower limit to the
uncertainty in the UV extinction can be calculated by assuming 
$\sigma[F(\lambda)_{\rm NEWSIPS}] = 0$ in
equation~\ref{eq_sig_fl}.  The resulting minimum uncertainties at all
UV wavelengths are 22, 19, 13, \& 13\% for AZV~18, 214, 398, \& 456,
respectively.  This illustrates the drawback of using moderately
reddened stars for extinction curve work.

  We fit each extinction curve with the Fitzpatrick \& Massa (1990,
hereafter \cite{fit90}) parameterization of the UV extinction curve.
This parameterization has a functional form with 3 terms.  The first
is a linear term ($c_1$ [y-intercept] and $c_2$ [slope]).  The second
is a Lorentzian-like ``Drude'' profile for the 2175 \AA\ bump ($c_3$
[strength], $x_o$ [bump center], and $\gamma$ [bump width]).  The
third is a curvature term for the far-UV ($x > 5.9~\micron^{-1}$,
$c_4$ [strength]).

\subsubsection{AZV 18 \label{sec_azv18}}

\begin{figure}[tbp]
\begin{center}
\plottwo{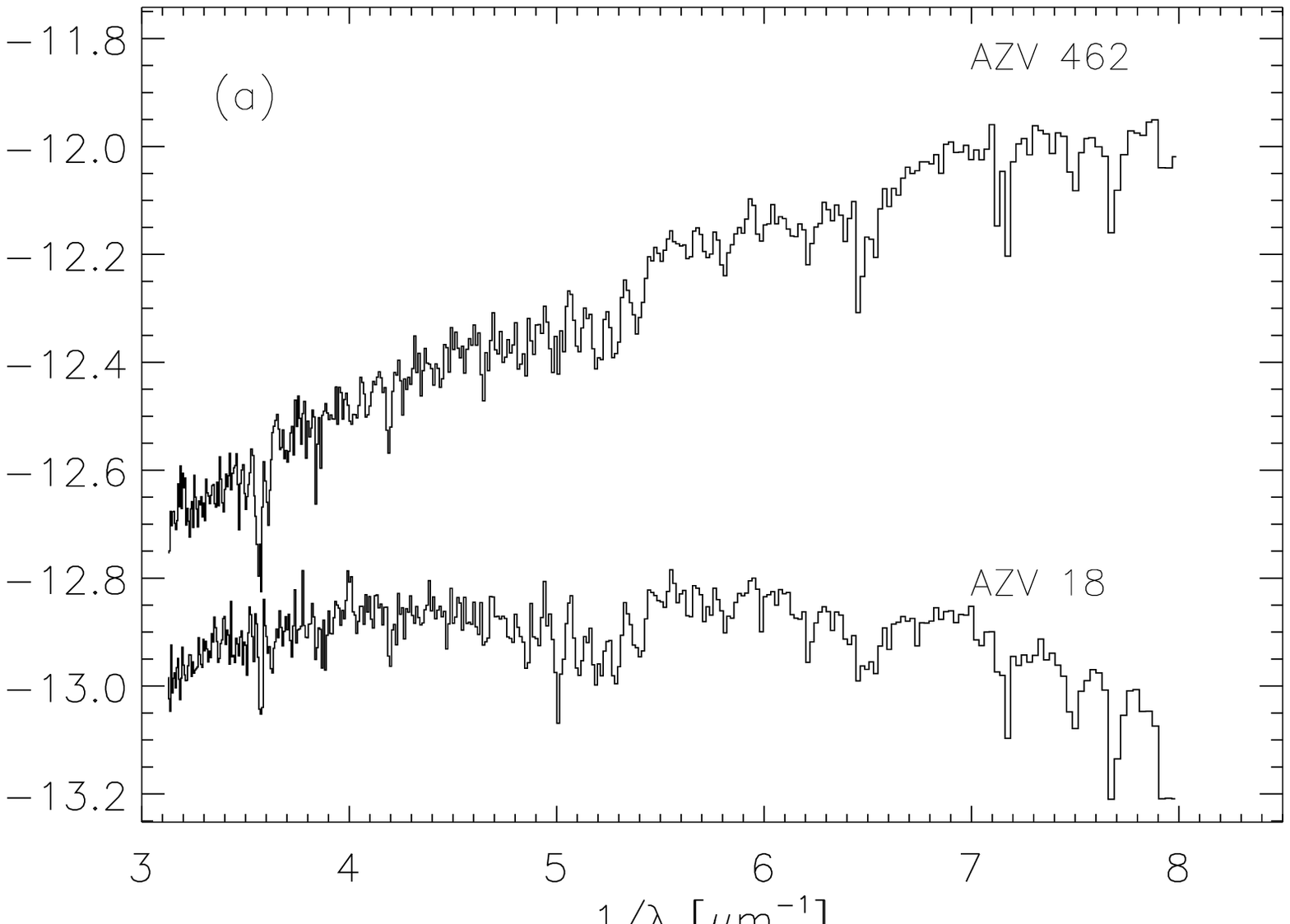}{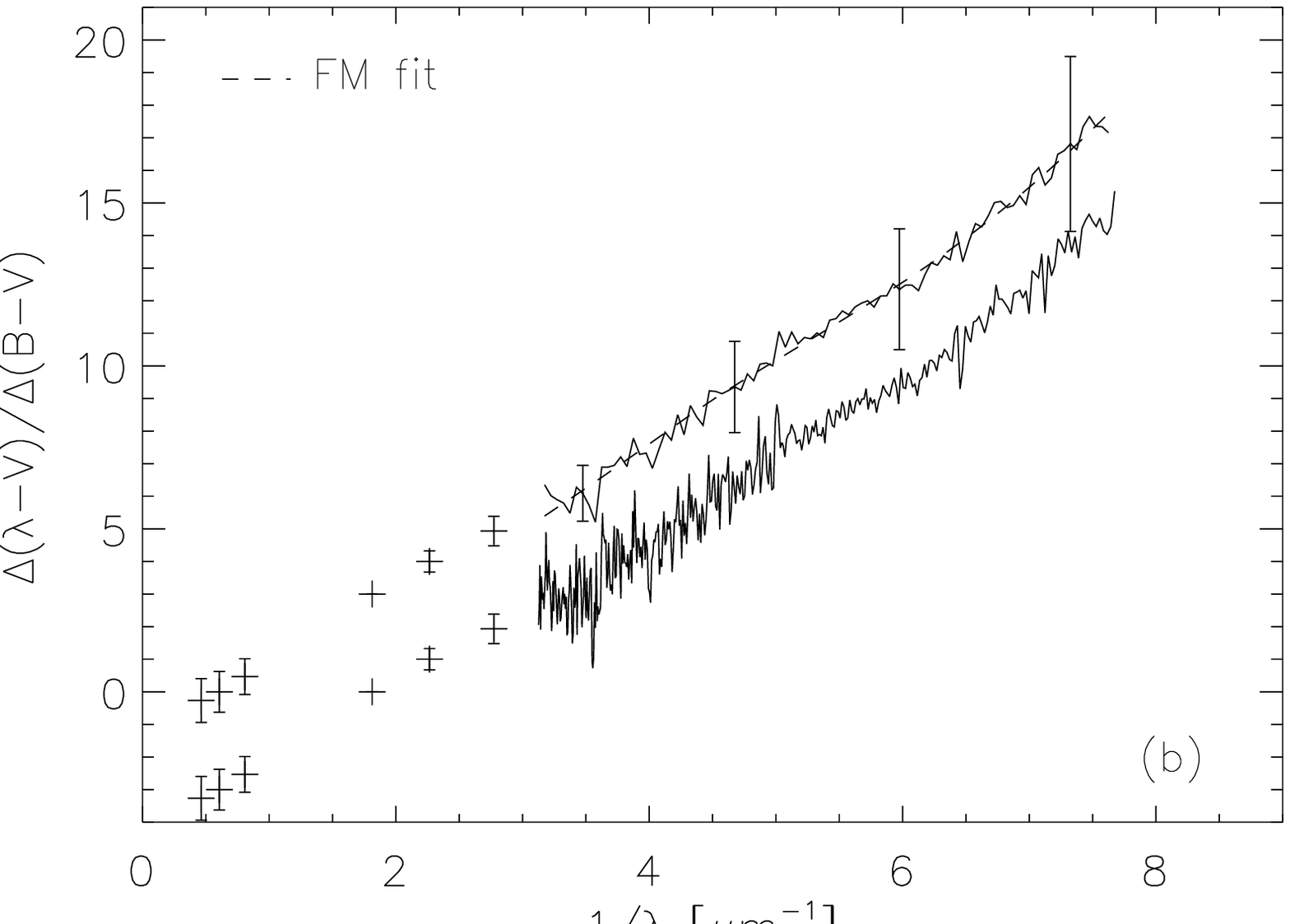}
\caption{The spectrum of reddened star AZV~18 and its comparison star
AZV~462 are plotted in (a).  The extinction curve derived for AZV~18
is displayed in (b) without any binning (lower) and with bins of 0.05
$\micron^{-1}$ and shifted by $\Delta E(\lambda) = 3$ (upper).  Plot
(b) also shows the FM fit for this extinction curve.
\label{fig_azv18_azv462}}
\end{center}
\end{figure}

  The extinction curve for AZV~18 has been determined previously
(\cite{pre84}; \cite{tho88}).  The comparison star which best
satisfies the Fitzpatrick criteria is AZV~462.  This star was one of
the three comparison stars used in Pr\'evot et al.\ (1984).  We only
used the extinction curve from the comparison star with the best
spectral match.  Averaging extinction curves made with multiple
comparison stars degrades the final extinction curve since the most
accurate extinction curve has then been averaged with less accurate
extinction curves.  Figure~\ref{fig_azv18_azv462}a displays the
spectra of AZV~18 and AZV~462.  The extinction curve for AZV~18 is
shown in Figure~\ref{fig_azv18_azv462}b.  The FM fit parameters are
tabulated in Table~\ref{table_FM_param}.

\subsubsection{AZV 214 \label{sec_azv214}}

  The UV extinction curve for the star AZV~214 has never been
calculated previously.  This is likely due to the presence of another
early-type star only $7\farcs 5$ away.  Both stars were included
within the {\it IUE} observing aperture.  In order to separate the
spectrum of AZV~214 from the nearby star, we used the MGEX and
NEWCALIB routines provided in the IUEIDL package to extract and
calibrate the spectrum for AZV~214 and the nearby star.  Neubig \&
Bruhweiler (1997) determined the UV spectral type using the combined
spectrum of the two stars.  This resulted in a UV spectral type too
late for AZV~214 since the nearby star had the effect of decreasing
the intensity of the spectral lines.  The optical and infrared
photometry of AZV~214 are likely unaffected by the nearby star as the
flux from this star, relative to AZV~214, is already small at
3000~\AA\ and decreasing to the red.  The nearby star is bluer than
AZV~214 and is probably an unreddened O type star.

\begin{figure}[tbp]
\begin{center}
\plottwo{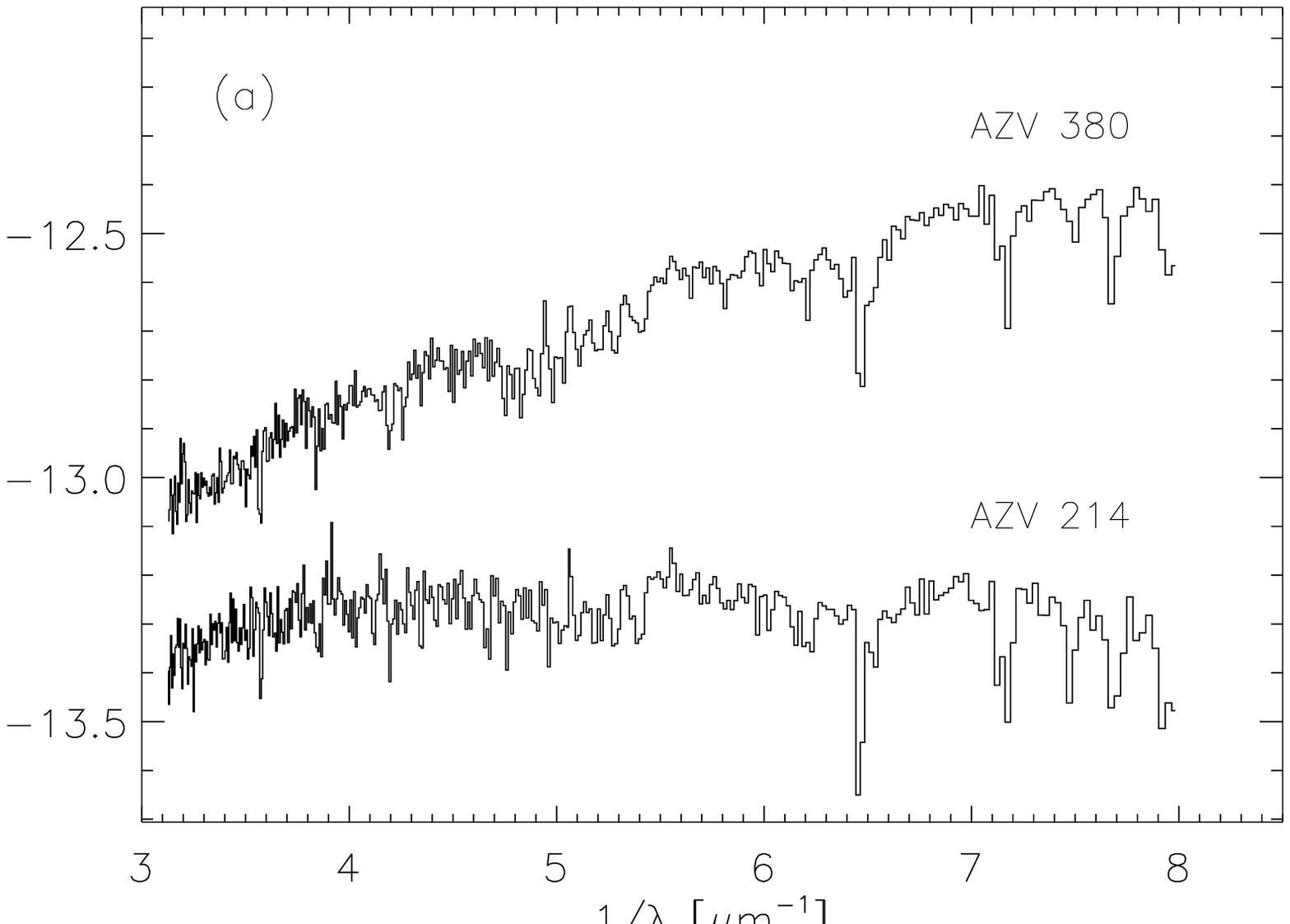}{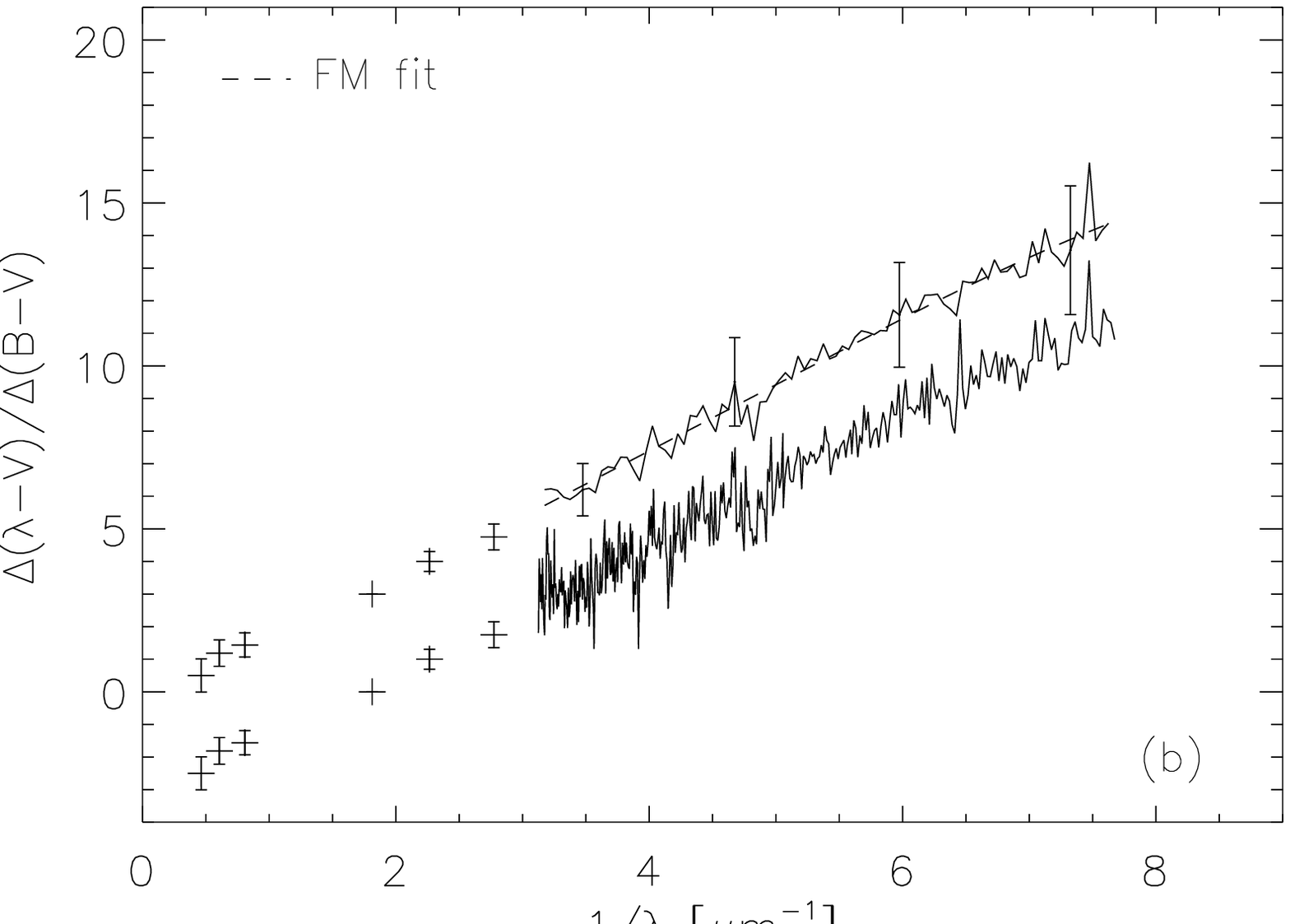}
\caption{The spectrum of reddened star AZV~214 and its comparison star
AZV~380 are plotted in (a).  The extinction curve derived for AZV~214
is displayed in (b) without any binning (lower) and with bins of 0.05
$\micron^{-1}$ and shifted by $\Delta E(\lambda) = 3$ (upper).  Plot
(b) also shows the FM fit for this extinction curve.
\label{fig_azv214_a_azv380}}
\end{center}
\end{figure}

   The comparison star which best satisfies the Fitzpatrick criteria
is AZV~380.  Figure~\ref{fig_azv214_a_azv380}a displays the spectra of
AZV~214 and AZV~380.  The extinction curve for AZV~214 is shown in
Figure~\ref{fig_azv214_a_azv380}b.  The FM fit parameters are
tabulated in Table~\ref{table_FM_param}.

\subsubsection{AZV 398 \label{sec_azv398}}

\begin{figure}[tbp]
\begin{center}
\plottwo{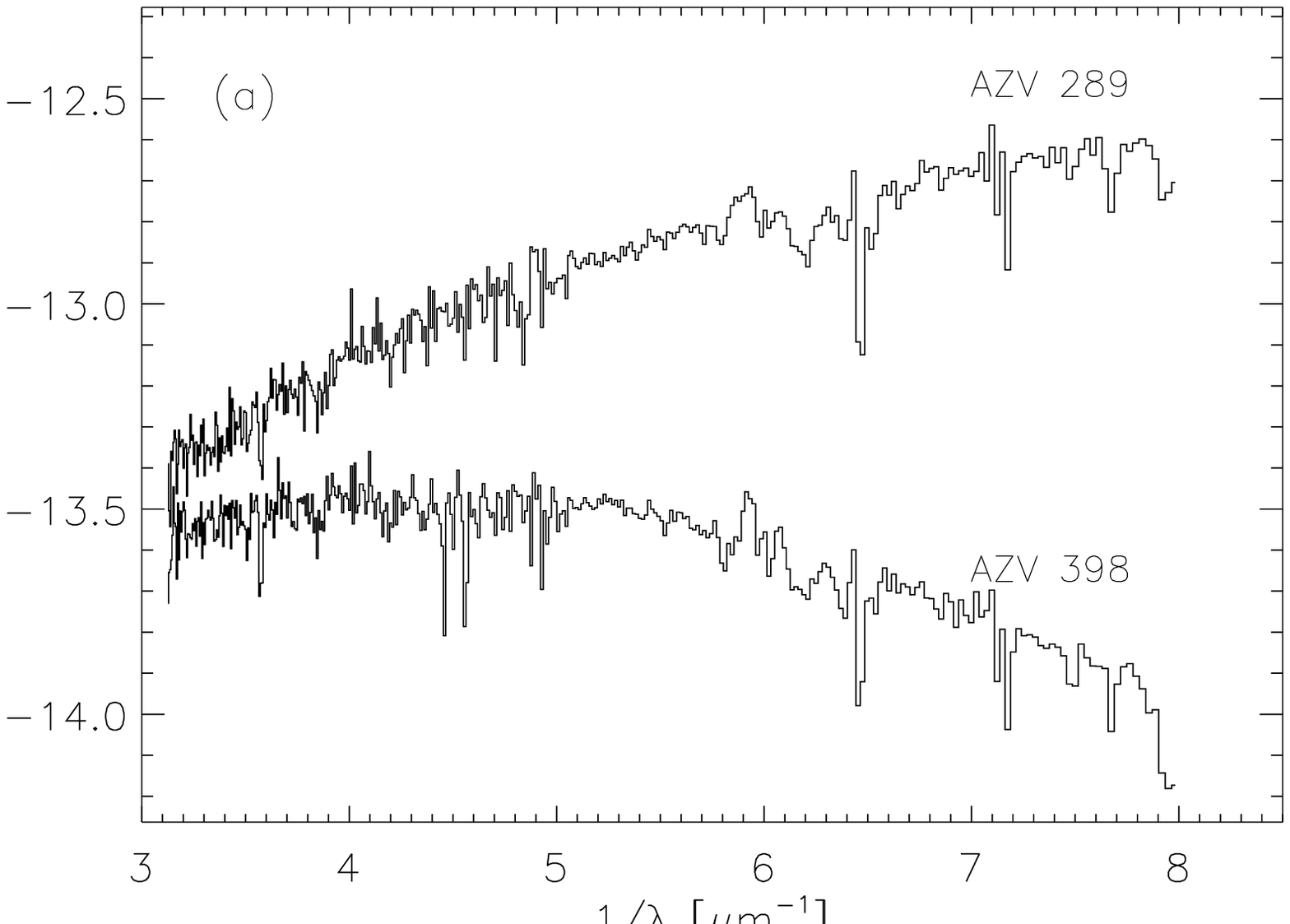}{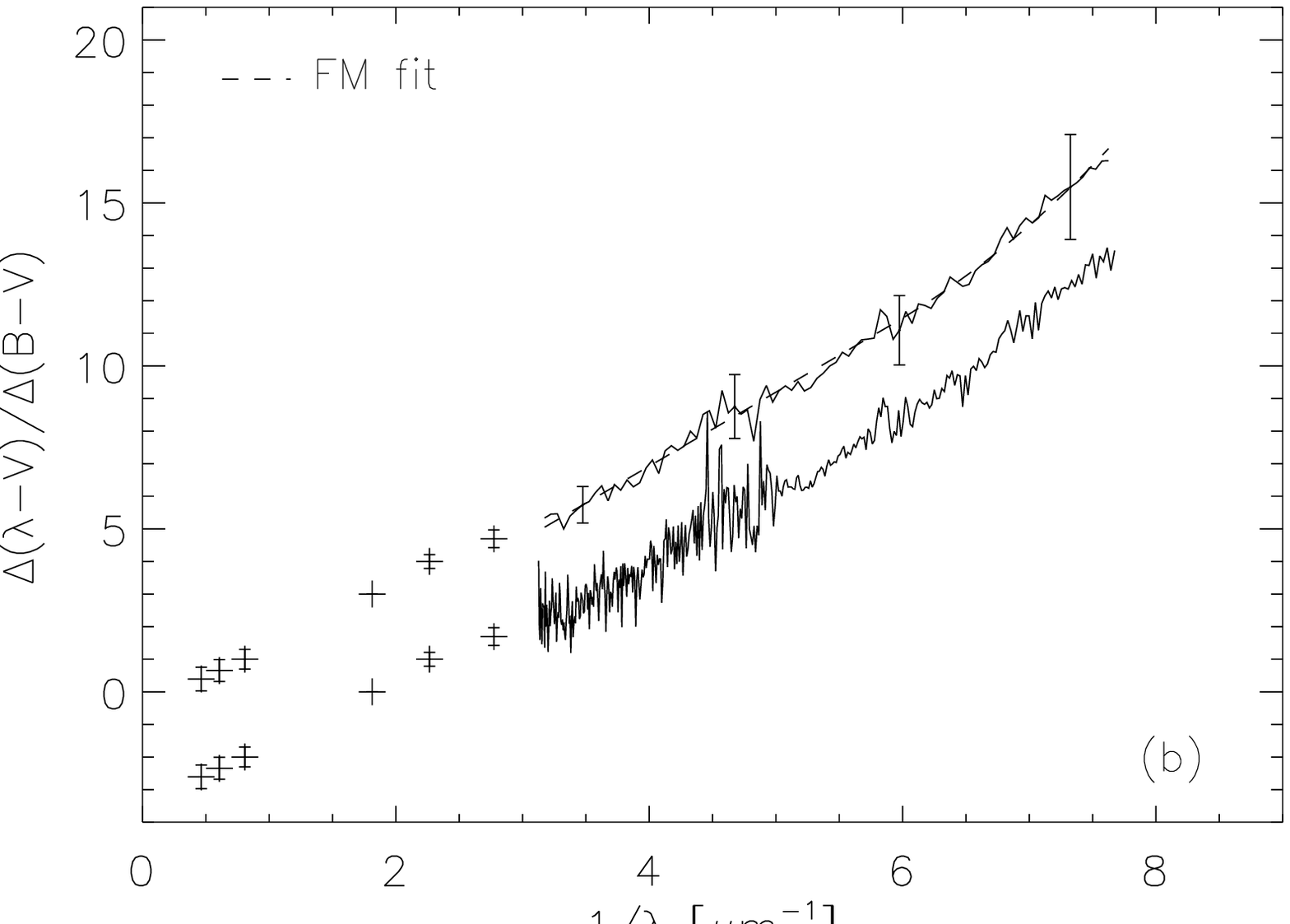}
\caption{The spectrum of reddened star AZV~398 and its comparison star
AZV~289 are plotted in (a).  The extinction curve derived for AZV~398
is displayed in (b) without any binning (lower) and with bins of 0.05
$\micron^{-1}$ and shifted by $\Delta E(\lambda) = 3$ (upper).  Plot
(b) also shows the FM fit for this extinction curve.
\label{fig_azv398_azv289}} 
\end{center}
\end{figure}

  The extinction curve for AZV~398 has been determined previously
(\cite{pre84}; \cite{rod97}).  The comparison star which best
satisfies the Fitzpatrick criteria is AZV~289.  This star was one of
the eight comparison stars used in Pr\'evot et al.\ (1984) and one of
the five stars used in Rodrigues et al.\ (1997).  Again, by using only
the comparison star with the best spectral match we were able to
determine a more accurate extinction curve.
Figure~\ref{fig_azv398_azv289}a displays the spectra of AZV~398 and
AZV~289.  The extinction curve for AZV~398 is shown in
Figure~\ref{fig_azv398_azv289}b.  The FM fit parameters are tabulated
in Table~\ref{table_FM_param}.

\subsubsection{AZV 456 \label{sec_azv456}}

\begin{figure}[tbp]
\begin{center}
\plottwo{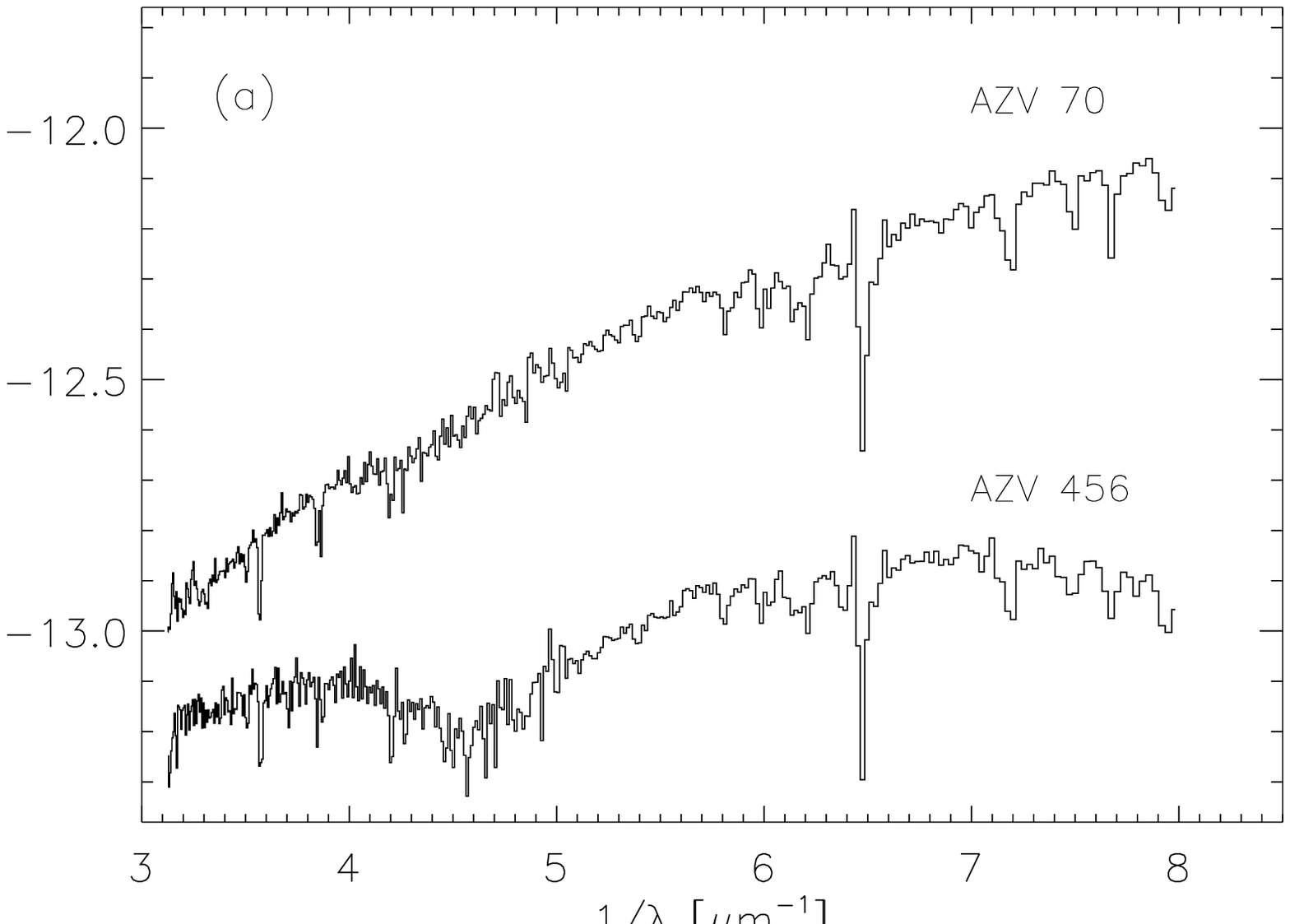}{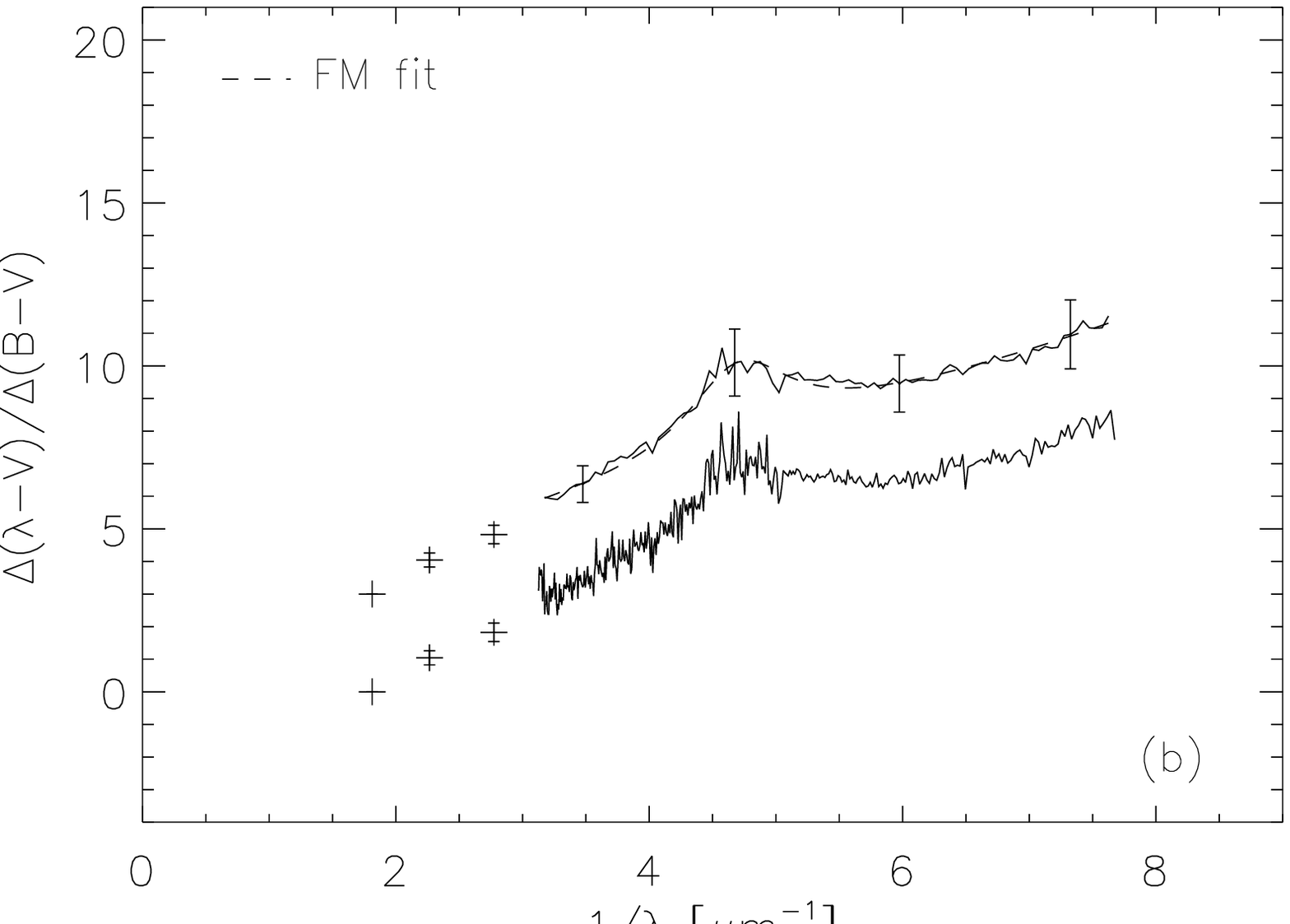}
\caption{The spectrum of reddened star AZV~456 and its comparison star
AZV~70 are plotted in (a).  The extinction curve derived for AZV~456
is displayed in (b) without any binning (lower) and with bins of 0.05
$\micron^{-1}$ and shifted by $\Delta E(\lambda) = 3$ (upper).  Plot
(b) also shows the FM fit for this extinction curve.
\label{fig_azv456_azv70}}
\end{center}
\end{figure}

  The star AZV~456 is the only star in the SMC to show the signature
of the 2175 \AA\ extinction bump in its spectrum
(Figure~\ref{fig_azv456_azv70}a).  We have chosen a different
comparison star than previous authors (\cite{leq82}; \cite{rod97})
specifically to remove mismatches in the \ion{C}{4} and \ion{Si}{4}
lines which are sensitive to both luminosity and temperature
(\cite{neu97}).  Figure~\ref{fig_azv456_azv70}a displays the spectra
of AZV~456 and AZV~70.  The extinction curve for AZV~456 is shown in
Figure~\ref{fig_azv456_azv70}b.  The FM fit parameters are tabulated
in Table~\ref{table_FM_param}.

\section{Discussion \label{sec_discussion}}

\begin{figure}[tbp]
\begin{center}
\plotone{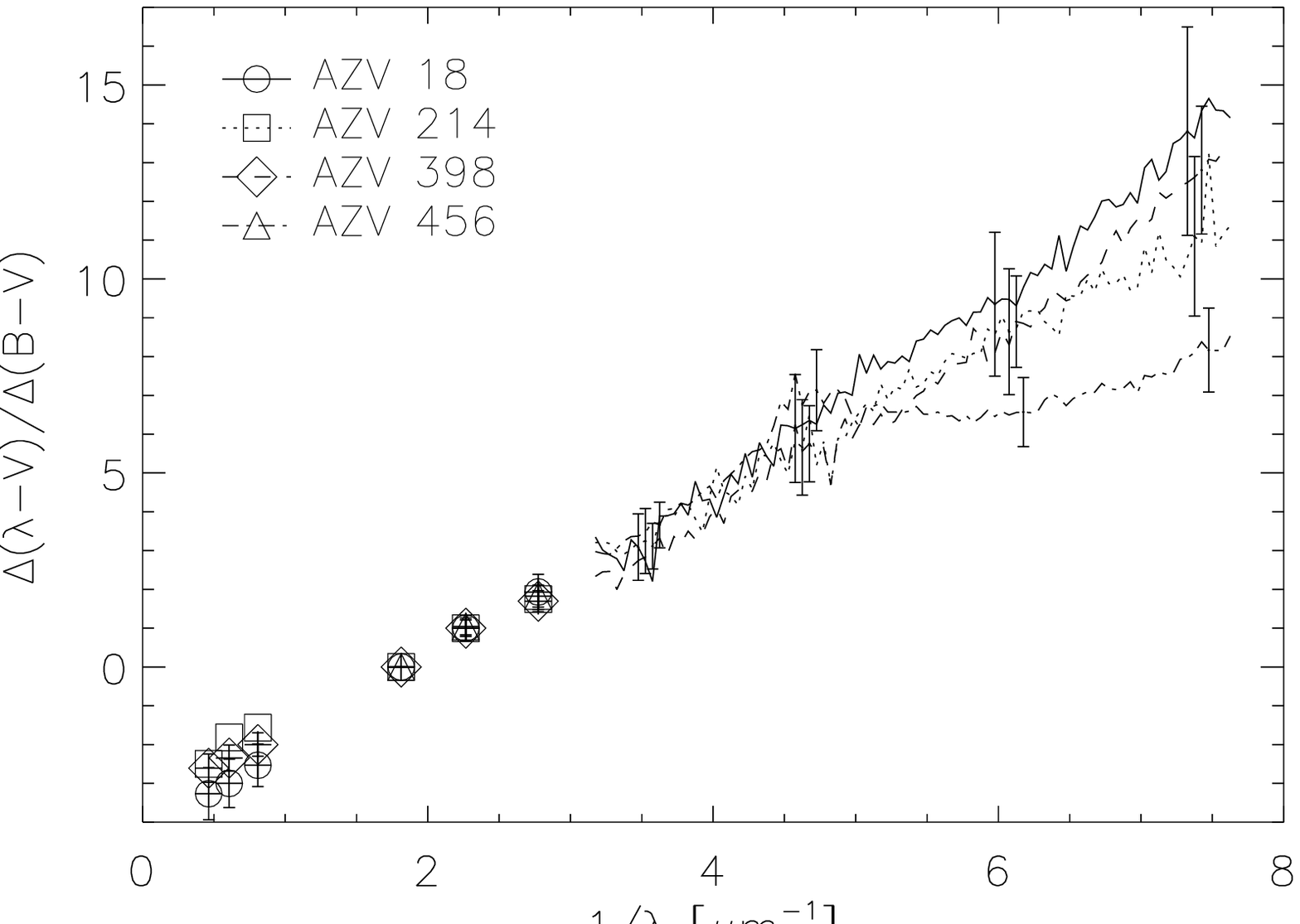}
\caption{The four extinction curves for AZV~18, 214, 398, \& 456 are
plotted. \label{fig_all_elv_ebv}}
\end{center}
\end{figure}

  Any discussion of the behavior of the dust in the SMC based on only
four extinction curves is obviously severely hampered by the small
sample size.  Yet, interesting trends can be seen even in this small
sample.  The four extinction curves are plotted together in
Figure~\ref{fig_all_elv_ebv} and their FM fit parameters are tabulated
in Table~\ref{table_FM_param}.  The extinction curves for AZV~18, 214,
\& 398 are very similar and are all roughly linear with
$\lambda^{-1}$.  The extinction curve for AZV~456 is much different as
it has a significant 2175~\AA\ bump and a weaker far-UV extinction.
The similarity of the extinction curves for AZV~18 \& 214 (lightly
reddened) to that of AZV~398 (more reddened) gives confidence that the
extinction curves for AZV~18 \& 214 are real.

\begin{deluxetable}{lcccccc}
\footnotesize
\tablewidth{0pt}
\tablecaption{FM fit parameters \label{table_FM_param}}
\tablehead{\colhead{AZV} & \colhead{$c_1$} &
           \colhead{$c_2$} & \colhead{$c_3$} & \colhead{$x_o$} &
           \colhead{$\gamma$} & \colhead{$c_4$}}
\startdata
18 & $-5.68 \pm 0.28$ & $2.53 \pm 0.05$ & $0.76 \pm 1.38$ &
  $4.56 \pm 0.23$ & $1.68 \pm 1.52$ & $0.60 \pm 0.26$ \nl
214 & $-3.72 \pm 0.29$ & $2.03 \pm 0.05$ & $0.09 \pm 3.03$ &
  $4.98 \pm 1.97$ & $5.12 \pm 19.24$ & $-0.20 \pm 0.26$ \nl
398 & $-5.16 \pm 0.29$ & $2.27 \pm 0.05$ & $0.14 \pm 11.29$ &
  $4.41 \pm 17.06$ & $-6.48 \pm 211.22$ & $0.80 \pm 0.28$ \nl
456 & $-0.96 \pm 0.09$ & $1.18 \pm 0.02$ & $2.57 \pm 0.22$ &
  $4.71 \pm 0.01$ & $1.00 \pm 0.05$ & $0.10 \pm 0.15$ \nl
\enddata
\end{deluxetable}

  Since the extinction curve for AZV~456 looks more like Milky Way
dust than SMC dust, one wonders if a large fraction of this extinction
is due to foreground Galactic dust.  Velocity resolved H I
measurements toward AZV~456 show that 90\% of the \ion{H}{1} along the
line-of-sight is located in the SMC (\cite{leq82}; \cite{mcg81},
1982).  Also, the star AZV~454, which is $1\farcm 5$ from AZV~456,
shows very little evidence of reddening.  These observations strongly
imply that most of the interstellar medium along the line of sight
toward AZV~456 lies in the SMC.

  In the Milky Way, most of the differences between extinction curves
can be explained by the variation in the single parameter $R_V =
A(V)/E(B-V)$.  This parameter measures the average dust grain size
with $R_V$ increasing with increasing average grain size
(\cite{car89}).  Following Bouchet et al.\ (1985), the $R_V$ values
for the four SMC extinction curves were calculated from
\begin{equation}
\label{eq_rv}
R_V = 1.10 \frac{\Delta (V - K)}{\Delta (B - V)}.
\end{equation}
The intrinsic colors of the reddened stars were assumed to be either
the colors of their respective comparison stars or the intrinsic
colors of Galactic stars of the same spectral types as tabulated in
Johnson (1966) for $(B-V)$, and Koornneef (1983), for $(V-K)$.  The
values of $R_V$ calculated both ways are tabulated in
Table~\ref{table_rv_values}.  The two $R_V$ values are equivalent
within the uncertainties and the adopted values are contained in
the last column of Table~\ref{table_rv_values}.  The four extinction
curves have roughly similar values of $R_V$.  If the SMC dust followed
a CCM-like relationship then AZV~456 would have the largest $R_V$.
This does not seem to be the case.

\begin{deluxetable}{lccc}
\tablewidth{0pt}
\tablecaption{$R_V$ Values \label{table_rv_values}}
\tablehead{\multicolumn{1}{c}{star} & \colhead{comparison} & 
           \colhead{Galactic} & \colhead{adopted} }
\startdata
AZV 18  & $3.60 \pm 0.73$ & $2.78 \pm 0.34$ & $3.60 \pm 0.73$ \nl
AZV 214 & $2.75 \pm 0.55$ & $2.36 \pm 0.21$ & $2.75 \pm 0.55$ \nl
AZV 398 & $2.87 \pm 0.40$ & $3.05 \pm 0.17$ & $2.87 \pm 0.40$ \nl
AZV 456 & \nodata         & $2.66 \pm 0.16$ & $2.66 \pm 0.16$ \nl
\enddata
\end{deluxetable}

\begin{figure}[tbp]
\begin{center}
\plotone{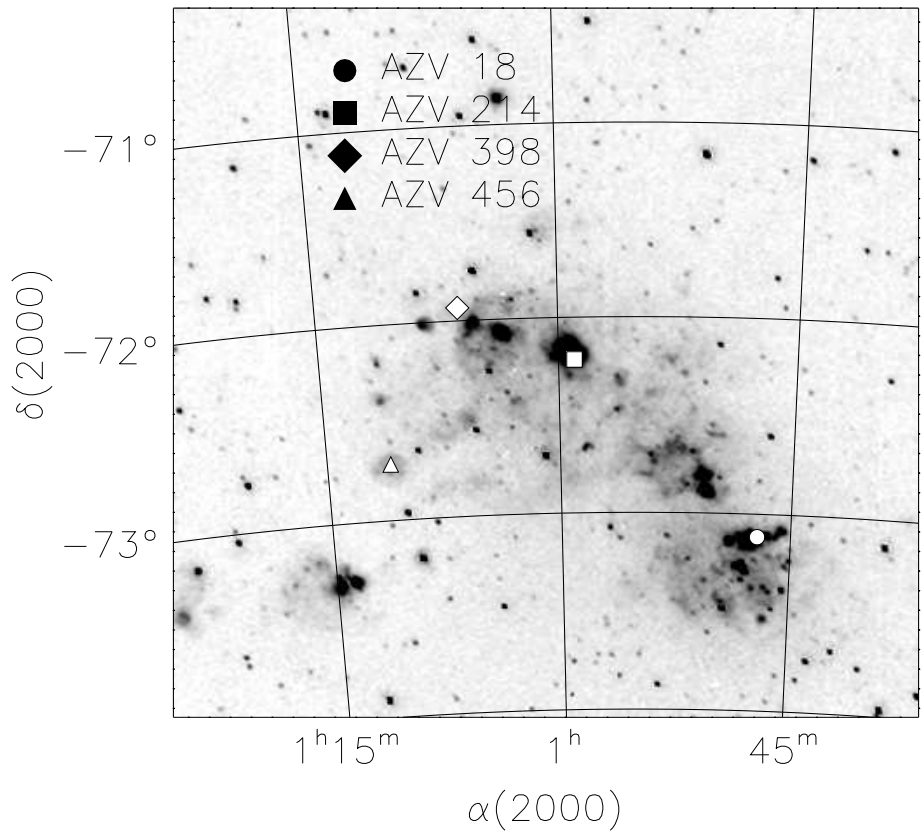}
\caption{The positions of the four reddened stars are plotted on a
H$\alpha$ image of the SMC (\cite{bot97}).  This image is displayed in
the North Celestial Pole projection (\cite{sta97}).  The image was
provided by G.\ Bothun and L.\ Staveley-Smith.
\label{fig_smc_Halpha}}
\end{center}
\end{figure}

  In order to investigate the dependence of the extinction in the SMC
on environment, we plotted the positions of the four stars on an
H$\alpha$ image of the SMC (Figure~\ref{fig_smc_Halpha}).  The
H$\alpha$ intensities trace star formation activity.  The
lines-of-sight toward all four stars are associated with the known
\ion{H}{2} regions (\cite{dav76}; \cite{cap96}).  The three stars with
roughly linear extinction curves (AZV~18, 214, \& 398) are located in
regions of high H$\alpha$ intensities (SMC Bar).  The one star
(AZV~456) with a more Galactic type extinction curve is also located
in an \ion{H}{2} region but one with much weaker star formation (SMC
Wing). The \ion{H}{2} region associated with the line-of-sight to 
AZV~214 is noteworthy because it is associated with the cluster
NGC~346 which is the most massive star formation region in the SMC.
NGC~346 contains 33 known O type stars (\cite{mas89}).  The H$\alpha$
flux from this cluster is 10\% of that seen from 30~Dor in the LMC
(\cite{mas89}).  The H$\alpha$ fluxes from the \ion{H}{2} regions
along the lines-of-sight toward AZV~18, 398, \& 456 are are 1, 1, and
0.1\%, respectively, of that seen for 30~Dor (\cite{ken86};
\cite{cap96}).  The dust along the AZV~456 sightline has likely been
exposed to a less harsh environment than the other three sightlines.
  
\begin{figure}[tbp]
\begin{center}
\plotone{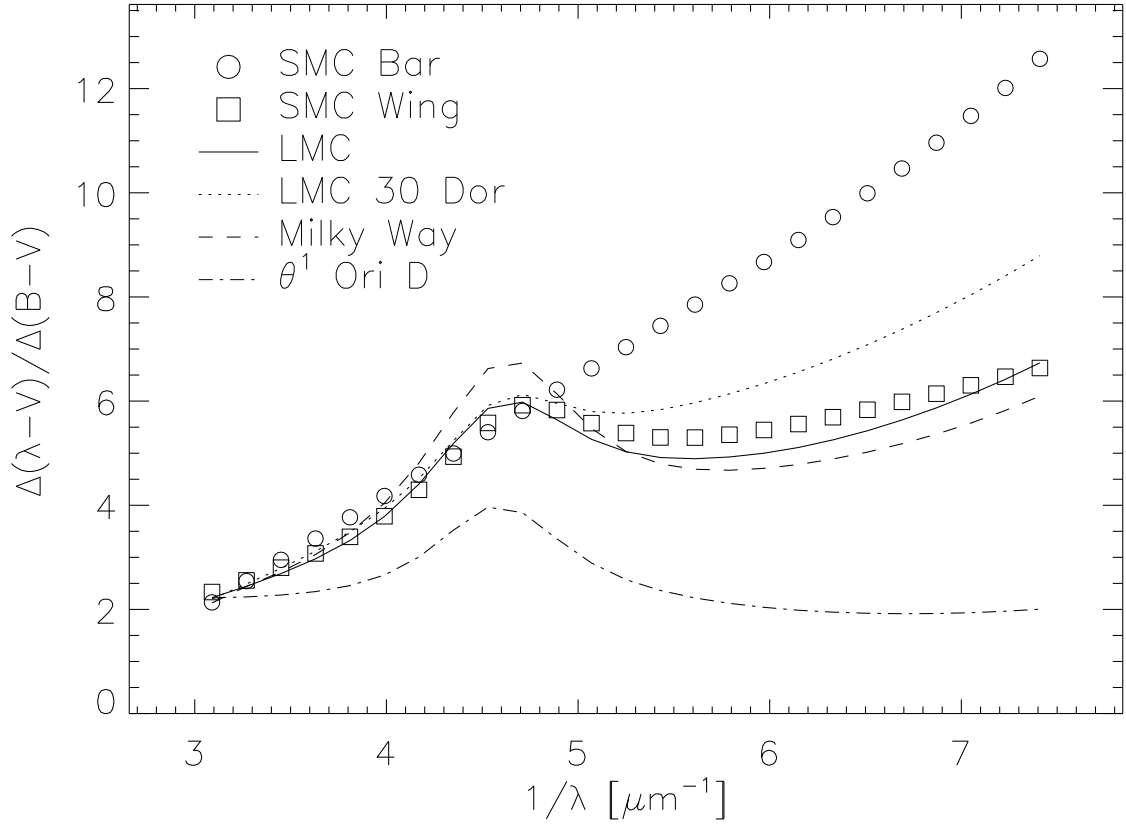}
\caption{The extinction curves for the SMC, LMC, and Milky Way are
plotted.  The curves plotted are those calculated from the FM fits
except for the Milky Way which was calculated from the CCM
relationship for an $R_V = 3.1$.  The extinction curve for the SMC Bar
is the $\Delta (B-V)$ weighted average of the curves for AZV~18, 214,
\& 398.  The SMC Wing and $\theta^1$~Ori~D extinction curves have been
multiplied by 0.83 and 1.3, respectively, to allow easier comparison
to the other four curves. \label{fig_all_ext}}
\end{center}
\end{figure}

  It is known that processing of Galactic dust near regions of active
star formation results in changes in the UV extinction curve
(\cite{mat92}).  A similar behavior is seen in the 30 Dor region in
the LMC (\cite{fit86}).  It is clear from this work that the harsh
radiation and shock environment associated with star formation regions
is modifying the dust in the SMC.  In all three galaxies, star
formation modifies nearby dust by altering the 2175~\AA\ bump and
increasing the strength of the far-UV extinction.  Yet, the extent of
the modification does {\em not} seem to be well correlated with the
level of star formation activity.  Figure~\ref{fig_all_ext} plots the
average extinction curves for the LMC (30~Dor and rest of LMC,
[\cite{fit86}]), the SMC (Bar and Wing), and the Milky Way ($R_V =
3.1$).  The extinction curve for $\theta^1$~Ori~D, a star in the Orion
\ion{H}{2} region, is also plotted.  The average extinction curve for
the SMC Bar is similar to the previous SMC average extinction curve
(\cite{pre84}).  From this figure, it is obvious that the extinction
in the SMC Bar is unlike that which has been found anywhere except
starburst galaxies (\cite{gor97}).  On the other hand, the extinction
in the SMC Wing is similar to that found in the LMC (excluding the
30~Dor region) and the Milky Way open cluster Trumpler~37 (not shown,
\cite{cla87}; \cite{fit90}).  The large difference between the
extinction curve in the Orion \ion{H}{2} region and the SMC Bar
implies that an Orion-like \ion{H}{2} region is not enough to produce
starburst-like dust.  Something more is needed.

  This raises the question: Why are the extinction properties of the
dust in the SMC Bar so much more extreme than that found in the 30~Dor
region of the LMC?  The largest star formation region in the SMC
(NGC~346) has only 10\% the activity of 30~Dor as measured by
H$\alpha$ (\cite{cap96}).  The extinction curve for AZV~456 shows that
dust, very similar to Galactic and LMC dust, exists in the SMC.
Therefore, there must be some significant environmental difference
between the LMC 30~Dor and SMC Bar regions which affects the extent to
which star formation activity can modify dust.  One known difference
between the LMC and SMC is metallicity.  The metallicities of the LMC
and SMC are 0.2 and 0.6 dex lower, respectively, than the that of the
local Galactic ISM (the ISM is 0.1 dex lower than solar).  However,
the relative abundances of the elements in the LMC and SMC are similar
to that found in the local interstellar medium (\cite{rus92}).
Metallicity is correlated with the dust-to-gas ratio in galaxies
(\cite{iss90}).  So, the amount of dust in the SMC is significantly
lower than that found in the LMC.  This could affect the ability of
dust grains in the SMC to shield themselves from the radiation and
shocks present near star formation regions.  This contrasts with the
finding that starburst galaxies all possess dust with an extinction
curve like that found in the SMC Bar even though they have
metallicities between 0.1 and 2.0 solar (\cite{gor97}).  It is
possible that the much higher level of star formation present in
starburst galaxies (10$\times$ that of 30~Dor) overwhelms other
environmental factors and always produces SMC Bar-like dust.  The
starburst galaxies were UV selected biasing the sample toward
intrinsically bright, nearby starbursts with at least one lightly
reddened starburst region.  Thus, the combination of a small column of
dust and the more intense star formation possibly accounts for the
presence of SMC-like dust in all starburst galaxies studied in Gordon
et al.\ (1997).

\section{Summary \label{sec_summary}}

  $\bullet$ We have greatly improved the UV extinction curves for
the SMC through improvements in the S/N of the IUE spectra and 
careful choices of reddened and comparison star pairs.

  $\bullet$ Four reddened SMC stars possess the Fitzpatrick criteria
needed for accurate extinction calculations.  Three of the four stars
possess a roughly linear (with $\lambda^{-1}$) extinction curve.  The
lines-of-sight toward these three stars (AZV~18, 214, \& 398) pass
through active star formation regions.  The fourth star (AZV~456) has
an extinction curve with a 2175~\AA\ bump and a weaker far-UV rise.
Its sightline also passes through a star formation region, but one
which is much less active.

  $\bullet$ Processing of dust near regions of star formation results in
variations in UV extinction. However, there is no simple correlation
between the strength of the variations and the amount of star
formation activity. Other parameters such as galaxy metallicity must
also play an important role.

  $\bullet$ As this work is based on only four sightlines, more
observations of reddened stars in the SMC are needed for to confirm
these results.  Of special need are observations outside the SMC Bar
to confirm the extinction curve of AZV~456.

\acknowledgements

  We thank G.\ Bothun and L.\ Staveley-Smith for giving us the
H$\alpha$ image.  This work was supported by NASA grant NAG5-3531.


\begin{thebibliography}{}
\bibitem[Ardeberg \& Maurice 1977]{ard77} Ardeberg, A.\ \& Maurice,
   E.\ 1977, \aaps, 30, 261
\bibitem[Azzopardi \& Vigneau 1979]{azz79} Azzopardi, M.\ \&
   Vigneau, J.\ 1979, \aaps, 35, 353
\bibitem[Azzopardi \& Vigneau 1982]{azz82} Azzopardi, M.\ \&
   Vigneau, J.\ 1982, \aaps, 50, 291
\bibitem[Azzopardi, Vigneau, \& Macquet 1975]{azz75} Azzopardi, M.,
   Vigneau, J., \& Macquet, M.\ 1975, \aaps, 22, 285
\bibitem[Bianchi et al.\ 1996]{bia96} Bianchi, L., Clayton, G.\ C.,
   Bohlin, R.\ C., Hutchings, J.\ B., \& Massey, P.\ 1996, \apj, 471,
   203
\bibitem[Bothun 1997]{bot97} Bothun, G.\ 1997, private communication
\bibitem[Bouchet et al.\ 1985]{bou85} Bouchet, P., Lequeux, J.,
   Maurice, E., Pr\'evot, L., \& Pr\'evot-Burnichon, M.\ L.\ 1985,
   \aap, 149, 330
\bibitem[Bouchet 1997]{bou97} Bouchet, P.\ 1997, private communication
\bibitem[Bromage \& Nandy 1983]{bro83} Bromage, G.\ E.\ \& Nandy, K.\
   1983, \mnras, 204, 29P
\bibitem[Calzetti et al.\ 1994]{cal94} Calzetti, D., Kinney, A.\ L.,
   \& Storchi-Bergmann, T.\ 1994, \apj, 429, 582
\bibitem[Caplan et al.\ 1996]{cap96} Caplan, J., Ye, T., Deharveng,
   L., Turtle, A.\ J., \& Kennicutt, R.\ C.\ 1996, \aap, 307, 403
\bibitem[CCM]{car89} Cardelli, J.\ A.,
   Clayton, G.\ C., \& Mathis, J.\ S.\ 1989, \apj, 345, 245 (CCM)
\bibitem[Cardelli, Sembach, \& Mathis 1992]{car92} Cardelli, J.\ A.,
   Sembach, K.\ R., \& Mathis, J.\ S.\ 1992, \aj, 104, 1916
\bibitem[Clayton \& Fitzpatrick 1987]{cla87} Clayton, G.\ C.\ \&
   Fitzpatrick, E.\ L.\ 1987, \aj, 92, 157
\bibitem[Clayton et al.\ 1996]{cla96} Clayton, G.\ C., et al.\ 1996,
   \apj, 460, 313
\bibitem[Clayton \& Martin 1985]{cla85} Clayton, G.\ C.\ \& Martin, P.\
   G.\ 1985, \apj, 288, 558
\bibitem[Davies, Elliott, \& Meaburn 1976]{dav76} Davies, R.\ D.,
   Elliott, K.\ H., \& Meaburn, J.\ 1976, \memras, 81, 89
\bibitem[Franx et al.\ 1997]{fra97} Franx, M., Illingworth, G.\ D.,
   Kelson, D.\ D., van Dokkum, P.\ G., \& Tran, K.-V.\ 1997, \apj,
   486, L75
\bibitem[Fitzpatrick 1985]{fit85} Fitzpatrick, E.\ L.\ 1985, \apj,
   299, 219
\bibitem[Fitzpatrick 1986]{fit86} Fitzpatrick, E.\ L.\ 1986, \aj, 92,
   1068
\bibitem[FM]{fit90} Fitzpatrick, E.\ L.\ \& Massa, D.\ 1990, \apjs,
   72, 163 (FM)
\bibitem[Garhart \& Nichols 1994]{gar94} Garhart, M.\ P.\ \& Nichols,
   J.\ S.\ 1994, in {\it IUE} Newsletter No.\ 55, 1
\bibitem[Garmany et al.\ 1987]{gar87} Garmany, C.\ D.,
   Conti, P.\ S., \& Massey, P.\ 1987, \aj, 93, 1070
\bibitem[Gordon et al.\ 1997]{gor97} Gordon, K.\ D.,
   Calzetti, D., \& Witt, A.\ N.\ 1997, \apj, in press (1 Oct)
\bibitem[Humphreys 1983]{hum83} Humphreys, R.\ 1983, \apj, 265, 176
\bibitem[Issa, MacLaren, \& Wolfendale 1990]{iss90} Issa, M.\ R.,
   MacLaren, I., \& Wolfendale, A.\ W.\ 1990, \aap, 236, 237
\bibitem[Johnson 1983]{joh83} Johnson, H.\ L.\ 1966, \araa, 4, 193
\bibitem[Kennicutt \& Hodge 1986]{ken86} Kennicutt, R.\ C., Jr.\ \&
  Hodge, P.\ W.\ 1986, 306, 130
\bibitem[Kinney et al.\ 1993]{kin93} Kinney, A.\ L., Bohlin, R.\ C.,
   Calzetti, D., Panagia, N., \& Wyse, R.\ F.\ G.\ 1993, \apjs, 86, 5
\bibitem[Koornneef 1983]{koo83} Koornneef, J.\ 1983, \aap, 128, 84
\bibitem[Lennon 1997]{len97} Lennon, D.\ J.\ 1997, \aap, 317, 871
\bibitem[Lequeux et al.\ 1982]{leq82} Lequeux, J., Maurice, E.,
   Pr\'evot-Burnichon, M.-L., Pr\'evot, L., Rocca-Volmerange, B.\
   1982, \aap, 113, L15
\bibitem[Lequeux et al.\ 1984]{leq84} Lequeux, J., Maurice, E.,
   Pr\'evot, L., Pr\'evot-Burnichon, M.-L., Rocca-Volmerange, B.\
   1984, in Structure and Evolution of the Magellanic Clouds, IAU
   Symp.\ 108, eds.\ S.\ van den Bergh \& K.\ S.\ de Boer (Dordrecht:
   Reidel) 405
\bibitem[Lowenthal et al.\ 1997]{low97} Lowenthal, J.\ D., et al.\
   1997, \apj, 481, 673
\bibitem[Madau, Pozzetti, \& Dickinson 1997]{mad97} Madau, P., Pozzetti,
   L., \& Dickinson, M.\ 1997, ApJ, in press
\bibitem[Massa, Savage, \& Fitzpatrick 1983]{mas83} Massa, D., Savage,
   B.\ D., \& Fitzpatrick, E.\ L.\ 1983, \apj, 266, 662
\bibitem[Massey et al.\ 1995]{mas95} Massey, P., Lang, C.\ C.,
   DeGioia-Eastwood, K., \& Garmany, C.\ D.\ 1995, \apj, 438, 188
\bibitem[Massey, Parker, \& Garmany 1989]{mas89} Massey, P., Parker,
   J.\ W., \& Garmany, C.\ D.\ 1989, \aj, 98, 1305
\bibitem[Mathis \& Cardelli 1992]{mat92} Mathis, J.\ S.\ \& Cardelli,
   J.\ A.\ 1992, \apj, 398, 610
\bibitem[McGee \& Newton 1981]{mcg81} McGee, R.\ X.\ \& Newton, L.\
   M.\ 1981, Proc.\ Astro.\ Soc.\ of Australia, 4, 189
\bibitem[McGee \& Newton 1982]{mcg82} McGee, R.\ X.\ \& Newton, L.\
   M.\ 1982, Proc.\ Astro.\ Soc.\ of Australia, 4, 308
\bibitem[Nandy et al.\ 1982]{nan82} Nandy, K., McLachlan, A.,
   Thompson, G.\ I., Morgan, D.\ H., Willis, A.\ J., Wilson, R.,
   Gondhalekar, P.\ M., \& Houziaux, L.\ 1982, \mnras, 201, 1P
\bibitem[Neubig \& Bruhweiler 1997]{neu97} Neubig, M.\ M.\ S.\ \&
   Bruhweiler, F.\ C.\ 1997, \aj, in press
\bibitem[Nichols et al.\ 1994]{nic94} Nichols, J.\ S., Garhart, M.\
   P., De La Pe\~na, M.\ D., \& Levay, K.\ L.\ 1994, New Spectral
   Image Processing System Information Manual: Low Dispersion Data -
   Version 1.0, {\it IUE} Newsletter No.\ 53
\bibitem[Pettini et al.\ 1997]{pet97} Pettini, M., Steidel, C.\ C.,
   Adelberger, K.\ L., Kellogg, M., Dickinson, M., Giavalisco, M.\
   1997, in ORIGINS, ASP Conference Series, eds.\ J.\ M.\ Shull, C.\
   E.\ Woodward, \& H.\ Thronson, in press (astro-ph/9708117)
\bibitem[Pr\`evot et al.\ 1984]{pre84} Pr\`evot, M.\ L., Lequeux, J.,
   Maurice, E., Pr\`evot, L., Rocca-Volmerange, B.\ 1984, \aap, 132,
   389
\bibitem[Rocca-Volmerange et al.\ 1981]{roc81} Rocca-Volmerange, B.,
   Pr\'evot, L., Ferlet, R., Lequeux, J., \& Pr\'evot-Burnichon, M.\
   L.\ 1981, \aap, 99, L5
\bibitem[Rodrigues et al.\ 1997]{rod97} Rodrigues, C.\ V.,
   Magalh\~aes, A.\ M., Coyne, G.\ V., \& Piirola, V.\ 1997, \apj,
   485, 618
\bibitem[Russell \& Dopita 1992]{rus92} Russell, S.\ C.\ \& Dopita,
   M.\ A.\ 1992, \apj, 384, 508
\bibitem[Sanduleak 1968]{san68} Sanduleak, N.\ 1968, \aj, 73, 246
\bibitem[Sanduleak 1969]{san69} Sanduleak, N.\ 1969, \aj, 74, 877
\bibitem[Staveley-Smith et al.\ 1997]{sta97} Staveley-Smith, L.,
   Sault, R.\ J., Hatzidimitriou, D., Kesteven, M.\ J., \& McConnell,
   D.\ 1997, \mnras, 289, 225
\bibitem[Steidel et al.\ 1996]{ste96} Steidel, C.\ C., Giavalisco, M.\
   Pettini, M., Dickinson, M., \& Adelberger, K.\ L.\ 1996, \apj, L17
\bibitem[Thompson et al.\ 1988]{tho88} Thompson, G.\ I., Nandy, K.,
   Morgan, D.\ H., \& Houziaux, L.\ 1988, \mnras, 230, 429
\bibitem[Trager et al.\ 1997]{tra97} Trager, S.\ C., Faber, S.\ M.,
   Dressler, A., \& Oemler, A., Jr.\ 1997, \apj, 485, 92
\bibitem[Walborn 1991]{wal91} Walborn, N.\ R.\ 1991, in Massive Stars
   in Starbursts, eds.\ C.\ Leitherer et al.\ (Cambridge: Cambridge
   Univ. Press), 145
\end{thebibliography}
\end{document}